\documentclass[10pt,conference,letterpaper]{IEEEtran}
\usepackage{times,amsmath,epsfig}
\usepackage{color}
\usepackage{algorithm}
\usepackage{algorithmic}
\title{Application-centric Resource Provisioning for Amazon EC2 Spot Instances}
\author{%

{Sunirmal Khatua{\small $~^{\#1}$}, Nandini Mukherjee{\small $~^{\$2}$} }%
\vspace{1.6mm}\\
\fontsize{10}{10}\selectfont\itshape

$^{\#}$\,Department of Computer Science and
Engineering, University of Calcutta,India\\
\fontsize{9}{9}\selectfont\ttfamily\upshape

$^{1}$\,skhatuacomp@caluniv.ac.in\\

\vspace{1.2mm}\\
\fontsize{10}{10}\selectfont\rmfamily\itshape

$^{\$}$\,Department of Computer Science and
Engineering, Jadavpur University,India\\
\fontsize{9}{9}\selectfont\ttfamily\upshape

$^{2}$\,nmukherjee@cse.jdvu.ac.in }

\begin{document}
\maketitle

\begin{abstract}
In late 2009, Amazon introduced spot instances to offer their
unused resources at lower cost with reduced reliability. Amazon's
spot instances allow customers to bid on unused Amazon EC2
capacity and run those instances for as long as their bid exceeds
the current spot price. The spot price changes periodically based
on supply and demand, and customers whose bids exceed it gain
access to the available spot instances. Customers may expect their
services at lower cost with spot instances compared to on-demand
or reserved. However the reliability is compromised since the
instances(IaaS) providing the service(SaaS) may become unavailable
at any time without any notice to the customer. Checkpointing and
migration schemes are of great use to cope with such situation. In
this paper we study various checkpointing schemes that can be used
with spot instances. Also we device some algorithms for
checkpointing scheme on top of application-centric resource provisioning framework that increase the
reliability while reducing the cost significantly.

\end{abstract}

\section{Introduction}
The era of cloud computing provides high utilization and high
flexibility of managing the computing resources. The elasticity
and on demand availability features of cloud computing ensure high
utilization of resources. Furthermore, resources can be availed
from templates that enforce standards so that resources can be
used with best management considerations without prior knowledge.
Therefore, flexibility is also high in cloud environment. The cloud
computing service models incorporate Infrastructure as a Service
(IaaS), Platform as a Service (PaaS) and Software as a Service
(SaaS). IaaS provides raw computing resources with different
capacity in the form of Virtual Machines(VM). Cloud service providers, like Google~\cite{google_cloud}, Amazon~\cite{amazon_cloud} provide these services and charge prices against these services from the clients. Among many such providers, Amazon defines the
capacity of resources in the form of 64 instance types \cite{instance_types} based on storage,
compute unit and I/O performance. The cost of these instance types
depends on the purchasing models defined by Amazon namely
on-demand, reserved and spot. On-Demand Instances let one pay for
compute capacity by the hour with no long-term commitments or
upfront payments. However with On-Demand Instances one may not
have access to the resources immediately. On the other hand, Reserved
Instances facilitate the client to make a low, one-time, upfront payment
for an instance, reserve it and get significant discount on hourly charge
over On-Demand Instances. Reserved Instances are always available for the
durations for which the clients reserve. In contrast with the above two
policies, where rates are fixed, Spot Instances provide the ability for
customers to purchase compute capacity with no upfront commitment
and at a variable hourly rates with a customer-defined upper
bound(bid) on the rate. Spot Instances are available only during
the time when the spot price is bellow the customer defined bid.

Thus spot instances make the resources unreliable in
nature and inappropriate for long running jobs like image
processing, gene sequence analysis etc. At the same time they
offer the opportunity to accomplish such jobs at a much lower cost
than on demand or reserved policies. Clearly checkpointing may be a good
option to make a tradeoff between the cost and reliability.
Checkpointing allows to store a snapshot of the current application state,
and later on, use it for restarting the execution at an opportunistic moment.

Various checkpointing techniques have been discussed in \cite{spot_ckpt} to
provide reliability with Amazon spot instances at lower cost. In this paper we study
some of these techniques and evaluate their performances. We also
investigate the effectiveness of application centric resource
provisioning framework \cite{khatua11} for actively monitoring the deployed
spot instances for an application and for taking necessary actions as the spot
intances become unavailable or the spot price changes. Finally
we propose and evaluate a novel checkpointing scheme for the application
centric resource provisioning framework.

The rest of the paper is organized as follows. A brief review of
the related works is presented in Section \ref{sec_priorwork}. An overview
of the application centric resource provisioning framework is given
in Section \ref{sec_framework}. The available resource provisioning options are described in Section \ref{sec_spot_characteristics}. Section \ref{sec_ckpt} discusses
the existing checkpointing schemes for spot instances while a proposed
checkpointing scheme for the application centric resource
provisioning framework is described in Section \ref{sec_ckpt_acc}. A simulated result for comparing the proposed checkpointing scheme
with existing ones is presented in Section
\ref{sec_implementation}. Finally we conclude with a direction of
future work in Sections \ref{sec_conclusion}.

\begin{figure*}
\centering
\includegraphics[width=7.0in]{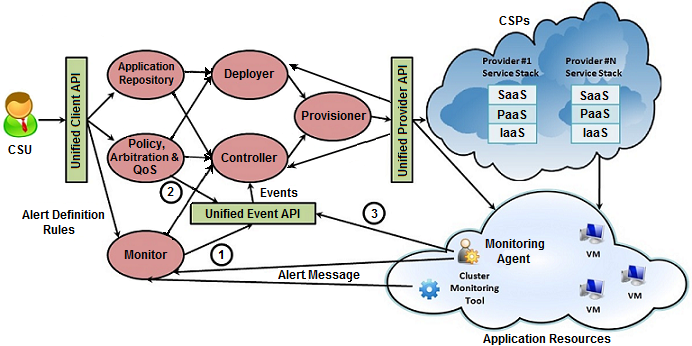}
\caption{The Application-centric resource provisioning framework}
\label{fig_architecture}
\end{figure*}

\section{Related Work}\label{sec_priorwork}
During the last couple of years, a lot of works \cite{khatua11}\cite{optimization_1}-\cite{optimization_3} concentrate on the cloud management aspect from the economic point of view. Most of them adapt a middleware based approach to optimize the resource requirement for a given cloud application. Paper \cite{khatua11} provides a novel framework for such a middleware. It identifies the key components of the middleware for auto deploying, auto scaling, providing robustness and availability of heterogeneous cloud applications. A model for optimal cloud resource scheduling based on stochastic integer programming technique is proposed in \cite{optimization_1}. A similar technique is also used in \cite{optimization_2} to optimize the resource requirement of a cloud application. This work tries to minimize the total provisioning cost by adjusting the tradeoff between the reservation and on-demand resource provisioning plans.

However a very few paper consider Amazon EC2 spot instances \cite{spot_instance} for providing economic benefit to cloud service users. S. Yi et. al. in their paper~\cite{spot_ckpt} not only consider the economic aspect of a cloud application but also the reliability of the application when running over the EC2 spot instances. They propose and simulate several checkpointing and migration schemes to reduce both job completion cost and job completion time on spot instances.

\section{Application-centric Resource Provisioning Framework}\label{sec_framework}
Traditional computing environment generally offers cloud services in
bottom up approach. Thus, the required infrastructure is set up
and then a specific platform is installed on top of that
infrastructure and finally applications are deployed on top of the
defined platform. Considering infrastructure to be fixed, the
variability increases as one goes up and the best combination of
platforms and applications are found to provide better utilization
of the infrastructure.  However, from a cloud user's point of view
the reverse is true. The user has an application and it is
required to find the best combination of SaaS, PaaS and IaaS to
provide better deployment at lower cost. Therefore, application
centric resource provisioning should adopt a top down approach
rather than a bottom up approach. Further, in such environment,
cost optimization techniques should be implemented
from the application's point of view, rather than the
infrastructure's point of view as followed in traditional
computing environments. Considering that the Cloud Service Provider (CSP) has
already optimized the use of the available physical resources,
the goal is to optimize the use of the virtual resources of Cloud Service Users (CSU)
for their deployed cloud applications.

Accordingly, we define an Application-centric resource provisioning framework \cite{khatua10} \cite{khatua11}
that will provide cost effective deployment of applications within a
common services platform. Each application is considered separately
with a specific combination of SaaS, PaaS and IaaS from a list of available providers as shown in
Figure \ref{fig_architecture}. A cloud application running on the framework requires to be formally defined
to deal with the open list of applications from a simple script to a complex n-tier system.
Thus, an application in the framework is defined with the following tuples \cite{khatua11}:

\begin{equation}
A = ( T, R, R_{m}, P, U, M )
\label{equ_a}
\end{equation}
\begin{tabbing}
\hspace{.2in}\=  where \=
$T$ is the set of tiers, (\{t\})\\
\> \> $R$ is the set of resources, (\{r\})\\
\> \> $R_{m}$ : $R \rightarrow T$\\
\> \> $P$ is the set of policies, (\{p\})\\
\> \> $U$ is the set of users (\{u\})\\
\> \> $M$ is the monitoring subsystem and
\end{tabbing}
\begin{equation}
M = ( E, W, E_{m}, W_{m} )
\label{equ_m}
\end{equation}
\begin{tabbing}
\hspace{.2in}\=  where \=
$E$ is the set of events, (\{e\})\\
\> \> $W$ is the set of workflows (\{w\})\\
\> \> $E_{m}$ : $E \rightarrow T~~|~~E \rightarrow R$\\
\> \> $W_{m} : W\rightarrow E$
\end{tabbing}

\begin{figure}
\centering
\includegraphics[width=3.5in]{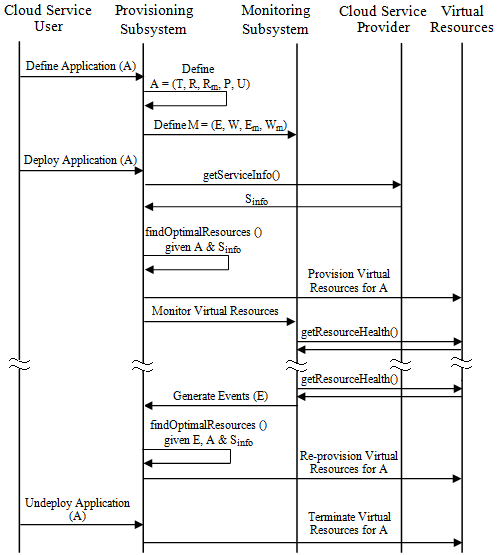}
\caption{Resource provisioning algorithm}
\label{seq_architecture}
\end{figure}

A brief description of functioning of the application centric resource provisioning framework is depicted in Figures \ref{fig_architecture} and \ref{seq_architecture}. The CSUs use a Unified Client API to define her application according to equations \ref{equ_a} and \ref{equ_m}. This unified definition of the applications are used by two important subsystems of the framework namely provisioning subsystem and monitoring subsystem as described below.

\subsection{Provisioning Subsystem}
The provisioning subsystem determines optimal provisioning of virtual resources for an application(A) satisfying the policies(P) specified for it. The application's required service level is stored in the policy(P). The provisioning subsystem queries various providers to get information about their offered services($S_{info}$). $S_{info}$ consists of provider id, service id, QoS id and the associated cost. The provisioning subsystem uses P(desired service level), $S_{info}$ and an optimization algorithm to find the optimal resource requirement for the application while maintaining the desired service level.

\subsection{Monitoring Subsystem}
\begin{figure}[h!]
\centering
\includegraphics[width=3.0in]{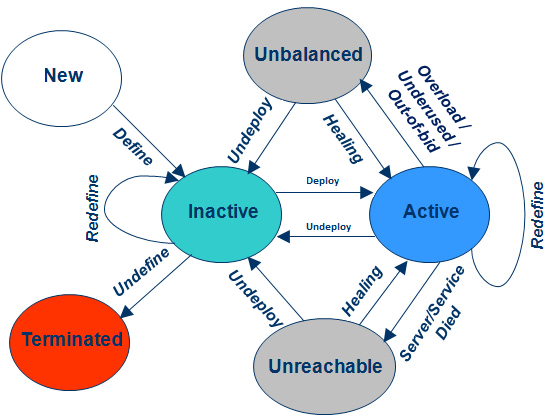}
\caption{States of an Application in application-centric cloud}
\label{fig_states}
\end{figure}

The monitoring subsystem implements a feedback system to inform the provisioning subsystem about the current state of the deployed application. The monitoring subsystem actively monitors the state of the deployed application and generates various events \cite{khatua10} to designate a change in the state.
In the proposed framework, an application can be in any of the six
defined states, namely New, Inactive, Active, Unbalanced,
Unreachable and Terminated (Figure \ref{fig_states}). Initially any application is in the
New state. Once such an application is mapped to various
modules according to the unified definition, the application enters
into the Inactive state. In the Inactive state the application is
composed (as specified by the unified definition), the required infrastructure is programmed (as determined by the optimization algorithm) and is ready
to be deployed within the cloud. The application is then deployed
and becomes ready to be accessed via the corresponding URL and its
state changes to the Active state. When in the Active state, the user pays for the cloud
resources. An application can be moved to the inactive state or to
the active state manually for fine tuning. If the application is
no longer required, the user can release the mapping from the
middleware and the application will be in the Terminated state.

Two other important states of the application are the Unbalanced
and Unreachable states. If the deployed application is overloaded
or underused based on certain threshold conditions, it reaches the
Unbalanced state. Similarly if any of the resource which is
deployed for the application fails, the application goes into the
Unreachable state. In these states, a workflow can be maintained
or generated in order to heal the situation and these actions send
back the application to the Active state. Figure~\ref{fig_states}
depicts the states of the application.

The monitoring subsystem uses different event
generation schemes for its proper operation. In \cite{khatua11}, five event generation
schemes are described. These are threshold based, prediction
based, request based, ping based and schedule based. The generated events carry necessary
information needed to reprovision the application resources to optimize or heal some undesired situation.
Once an event is generated, the monitoring subsystem sends the event to
the provisioning subsystem. Once an event(E) is received, the  provisioning subsystem analyzes the event and uses E, P, $S_{info}$ and an optimization algorithm
for reprovisioning the application onto appropriate resources.

\section{Resource Provisioning in Amazon EC2 Cloud} \label{sec_spot_characteristics}
In this paper multiple providers of application centric resource provisioning are not considered. Rather, we consider various resource provisioning options available from Amazon EC2 public provider only. Amazon sells their resources in the form of on-demand, reserved
and spot instances.

On-demand resources can be used without any upfront payment and just paying as much as the client use on a hourly basis.
However request for on-demand instances may not be met immediately due to unavailability of Amazon EC2 resources. Thus for a
long term and time critical application it is required to opt for reserved
instances. With reserved instances required resources can be reserved
with some upfront payment and access to the
reserved instances can be made whenever the client needs. Amazon also provides
competitive discounts on the hourly charge for the reserved
instances. The third category of the instances, i.e. spot
instances, allow the user to use Amazon's unused resources at lower cost
compared to on-demand and reserved instances if available. The prices of spot
instances, called spot price, depends on the demand and supply of
the specific instance type at a specific availability zone. Users
need to define the bid (the maximum cost he is willing to pay per instance) for a specific
instance type at a specific availability zone and the spot
instance request will be granted if the current spot price is less
than the bid defined by the user.\\\\
\textbf{\textit{Characteristics of Amazon EC2 Spot Instances:}}

The variable price of spot instances makes them an important consideration for optimizing resource requirement for an application. However, their volatile nature makes them inherently unreliable and hence the optimization algorithms become more challenging than the other instances.

\begin{figure}[h!]
\centering
\includegraphics[width=3.5in]{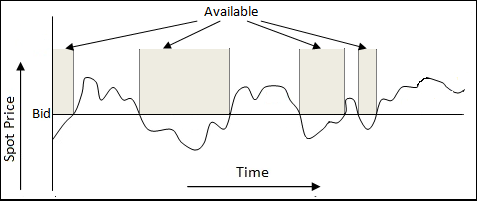}
\caption{Availability of a spot instance}
\label{fig_availability}
\end{figure}

Various characteristics of Amazon EC2 spot instances \cite{spot_instance} are summarized below:

\begin{itemize}
\item Spot instances are available when the user's bid exceeds the
current spot price (Fig.~\ref{fig_availability}).
\item Spot instances are terminated (becomes
unavailable) without any notification to the user whenever the
current spot price exceeds the user's bid.
\item The price per instance-hour for
a spot instance is set at the beginning of each instance-hour. Any
change to the spot price will not be reflected until the next
instance-hour begins.
\item Amazon will not charge the last
partial hour if the spot instance is terminated due to out-of-bid
situation. However Amazon will charge the full hour if the user
terminate the instance forcefully.
\item Amazon provides the history of spot prices of a spot instance at a specific availability zone for the last 3 months free of cost.
\end{itemize}

\section{Checkpointing Schemes for Spot Instances}\label{sec_ckpt}

The characteristics of spot instances make them appealing
for long running jobs with divisible workloads~\cite{divisible_workload}. Various existing
checkpointing schemes can be adopted for saving the completed
tasks and resuming the remaining tasks as and when the spot
instances become available.\\\\
\textit{\textbf{Existing Checkpointing Schemes:}}

The checkpointing schemes proposed in \cite{spot_ckpt} are briefly described below:

1. No Checkpointing~(NONE): Checkpoints are not taken and all the
tasks for a job are required to be repeated after every out-of-bid events.

2. Optimal Checkpointing~(OPT): Checkpoints are taken just prior to
the out-of-bid events. Clearly it will save the maximum number of
tasks out of each available interval for a given instance type and
a user's bid.

3. Hourly Checkpointing~(HOUR): Checkpoints are taken just prior to
the beginning of next instance hour. Since Amazon is not charging
any partial hour, this scheme will save as much tasks as the user
is paying.

4. Rising edge-driven Checkpointing~(EDGE): Checkpoints are taken
after every increase~(rising edge) of the current spot price.

5. Adaptive Checkpointing~(ADAPT): Checkpoints are taken or skipped
at regular intervals based on the expected recovery time for
taking or skipping the checkpoint. It will take a checkpoint if the
expected recovery time is higher for skipping the checkpoint. The expected recovery
time is calculated using a probability density function of expected out-of-bid events.
Such a probability density function is determined from the history of spot prices and the user defined bid.

Out of the above five checkpointing schemes NONE and OPT provide
two extreme results without any practical value. They are used to
provide comparative study of the other realistic checkpointing schemes.

\section{A Novel Checkpointing Scheme for Application-Centric Resource Provisioning}\label{sec_ckpt_acc}
In this section we propose a novel checkpointing scheme for spot instances on top of application-centric resource provisioning framework. For the purpose we devise a new event generation scheme that deals with spot instances. The new checkpointing scheme is targeted to achieve performance comparative to OPT checkpointing scheme described above. Before describing the scheme, we introduce a modified event generation scheme for our application-centric resource provisioning framework.

\subsection{Event Generation Scheme for Spot Instances}\label{sec_event_generation}
The event generation schemes proposed in \cite{khatua11} is extended to include new events that support spot instances. As discussed in Section \ref{sec_spot_characteristics}, the availability of spot instances depends on the current spot price and the user defined bid. Also spot instances become unavailable without prior notification to the clients that makes them inherently unreliable. The reliability can be increased by taking checkpoints (saving completed tasks) during the available periods. However, the time of taking checkpoints affects the reliability as well as job completion time and cost.

\begin{figure}[h!]
\centering
\includegraphics[width=3.5in]{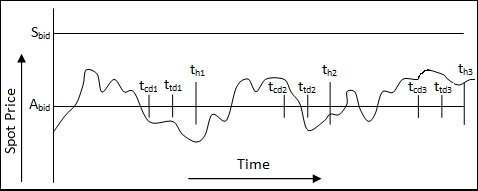}
\caption{Decision Points for Event Generation}
\label{fig_dp}
\end{figure}

Accordingly, in this paper we propose a new event generation scheme to handle spot instances. Three events are proposed, namely $E_{ckpt}$, $E_{terminate}$ and $E_{launch}$. $E_{ckpt}$ is used for taking checkpoint, $E_{terminate}$ is used to terminate a spot instance forcefully and $E_{launch}$ is used to relaunch a previously terminated spot instance. We define two bid
values for the purpose - one for the application($A_{bid}$) and other for the spot
instance($S_{bid}$). $S_{bid}$ is sufficiently large and is used in the
request for spot instance. Clearly, the value is maintained at such a high level, that Amazon will never terminate
the spot instances due to out-of-bid situation. On the other hand, $A_{bid}$ is user
defined bid for the application and is stored in the monitoring subsystem as part of the event definition(E) of the application. The Monitor module actively monitors the current spot price and generates the two events, $E_{ckpt}$ and $E_{terminate}$, for the Controller
module. On the basis of these two events, the Controller module either takes a checkpoint or terminate the corresponding spot instance respectively. However to increase the performance, the Controller module will query the current spot price only at specific points of time called decision points. Since the cost of spot instance is not changed during an instance hour and is fixed at the beginning of that instance hour, the decision points should be relative to the beginning of an instance hour. Accordingly we define two decision points just prior to each hour boundary as follows:
\begin{equation}
t_{cd} = t_{h} - t_{c} - t_{w}
\label{equation_t_cd}
\end{equation}
\begin{equation}
t_{td} = t_{h} - t_{w}
\label{equation_t_td}
\end{equation}
where $t_{cd}$ and $t_{td}$ are the decision points for
checkpointing and terminating a spot instance. $t_{h}$ is an hour
boundary, $t_{c}$ is the time needed to take a checkpoint and
$t_{w}$ is the waiting time to get the current spot price. The Monitor module will generate $E_{ckpt}$ at $t_{cd}$ if the current spot price exceeds $A_{bid}$ and will generate $E_{terminate}$ at $t_{td}$ if the current spot price is still above the $A_{bid}$. It will generate $E_{launch}$ at the start of each available period of a spot instance with respect to $A_{bid}$. This event generation scheme is illustrated in Figure~\ref{fig_dp}. It will generate neither $E_{ckpt}$ nor $E_{terminate}$ for the hour boundary $t_{h1}$. It will generate $E_{ckpt}$ but not $E_{terminate}$ for the hour boundary $t_{h2}$.
For the hour boundary $t_{h3}$, it will generate both $E_{ckpt}$ and $E_{terminate}$ since the user will have to pay
above $A_{bid}$ for the next hour.

\subsection{The Application-Centric Checkpointing Scheme}\label{subsection_acc}
\begin{figure}[h!]
\centering
\includegraphics[width=3.5in]{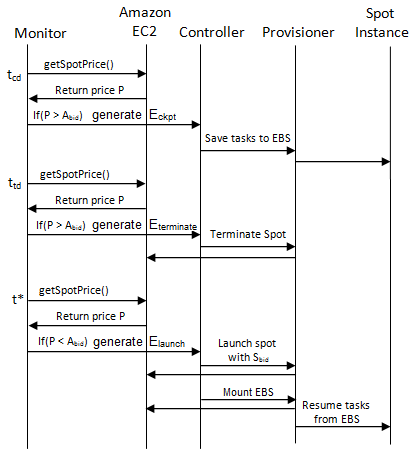}
\caption{Application Centric Checkpointing Scheme} \label{seq_acc}
\end{figure}

In this section, we propose a checkpointing scheme on top of the application
centric resource provisioning framework, called Application
Centric Checkpointing(ACC). ACC is based on the event generation scheme discussed in the previous subsection and is described by the sequence diagram shown in Figure~\ref{seq_acc}.

The following unified definition can be used for an application with divisible workloads to be run on spot:
\begin{equation}
A = ( T, R, R_{m}, P, U, M )
\label{equ_a1}
\end{equation}
\begin{tabbing}
\hspace{.2in}\=  where \=
$T$ = \{$t_{1}$\}\\
\> \> $R$ = \{$r_{1},r_{2}$\}\\
\> \>~~~~~~$r_{1}$.provider = ec2, $r_{1}$.type = spot instance,\\
\> \>~~~~~~~~~~~~~~~~~~~~$r_{1}$.size = $<instance\_type>$\\
\> \>~~~~~~$r_{2}$.provider = ec2, $r_{2}$.type = EBS,\\
\> \>~~~~~~~~~~~~~~~~~~~~$r_{2}$.size = 1GB\\
\> \> $R_{m}$ = \{ $r_{1} \rightarrow t_{1}$, $r_{2} \rightarrow t_{1}$ \}\\
\> \> $P$ = \{ $sla$ \}\\
\end{tabbing}
\begin{equation}
M = ( E, W, E_{m}, W_{m} )
\label{equ_m1}
\end{equation}
\begin{tabbing}
\hspace{.2in}\=  where \=
$E$ = \{$E_{ckpt}$,~$E_{terminate}$,~$E_{launch}$\},\\
\> \> ~~~~~~~~$threshold~for~all~events = <A_{bid}>$,\\
\> \> ~~~~~~~~$E_{launch}.bid = <S_{bid}>$\\
\> \> $W$ = \= \{$W_{start}$,~$W_{ckpt}$,~$W_{terminate}$,~$W_{launch}$\}\\
\> \> \> ~~$W_{start}$ = \{ \= Launch spot; \\
\> \> \> \> Mount EBS; \\
\> \> \> \> Copy job to EBS; \\
\> \> \> \> Start job \},\\
\> \> \> ~~$W_{ckpt}$ = \{Save results to EBS\},\\
\> \> \> ~~$W_{terminate}$ = \{Terminate spot\} \& \\
\> \> \> ~~$W_{launch}$ = \{ \= Launch spot; \\
\> \> \> \> Mount EBS; \\
\> \> \> \> Resume tasks \},\\
\> \> $E_{m}$ = \{$E_{ckpt} \rightarrow r_{1},~E_{terminate} \rightarrow r_{1}$,\\
\> \> \> ~~~$E_{launch} \rightarrow r_{1} \})$\\
\> \> $W_{m}$ = \{$W_{ckpt} \rightarrow E_{ckpt}$,\\
\> \> \> ~~~$W_{terminate} \rightarrow E_{terminate},$\\
\> \> \> ~~~$W_{launch} \rightarrow E_{launch}$\}\\
\end{tabbing}

The Elastic Block Storage (EBS) \cite{ebs} is used to save the completed tasks
during checkpoint. The parameters $instance\_type$, $A_{bid}$ and $S_{bid}$ can be set either manually by the end user or by some optimization or greedy algorithms. The provisioning subsystem~(Deployer module) can use the following simple greedy strategy for choosing $A_{bid}$ and $instance\_type$:

\begin{algorithm}
\caption{Determine $A_{bid}$ \& $instance\_type$}
\begin{algorithmic}
\item[1.] Retrieve $S_{info}$ from Amazon EC2. \\/*\texttt{~$S_{info}$ carries availability zone, spot instance type and history of spot price.}*/
\item[2.] Find the list of instance types that meet the required service level agrement(sla) specified in P. \\/*\texttt{~The list is denoted by L.}*/
    \item[3.] Calculate application bid as
        \begin{equation}
        A_{bid} = \min C_i, ~\forall i \in L
        \label{equation_a_bid}
        \end{equation}
         \\/*\texttt{~$C_i$ is the corresponding on demand instance's cost per hour for the instance type i.}*/
    \item[4.] For each instance type $i \in L$
    \begin{enumerate}
       \item[4.1]  Calculate Expected Execution Time~(EET) for a job of length `w' when executed in a spot instance of instance type `i' with a bid of value $A_{bid}$.
            \begin{equation}
            EET_i = \frac{w\sum_{k=w}^{\infty }f_i(k)+\sum_{k=0}^{w-1}(k+r)f_i(k)}{1-\sum_{k=0}^{w-1}f_i(k)}
            \label{equation_eet}
            \end{equation}
            \\/*\texttt{~$f_i$(t) is the probability density function of the spot instance type i's  failure for out-of-bid. The $f_i$(t) is calculated from the spot instance type i's history of spot price and $A_{bid}$.}*/
    \end{enumerate}
    \item[5.]  Choose $instance\_type$ = i $\mid$ $EET_i$ is minimum.
\end{algorithmic}
\end{algorithm}

After determining the parameters $A_{bid}$ \& $instance\_type$, the Deployer module starts $W_{start}$ workflow. The $W_{start}$ workflow launches a spot instance as per the specification of the resource $r_1$ and an EBS volume as per the specification of the resource $r_2$. The workflow then mounts the EBS volume to the spot instance, copy the job from the application repository to the EBS and starts the job.

Once the application is deployed, EC2 starts charging for the resources. The monitoring subsystem~(Monitor module) calculates $t_{cd}$ and $t_{td}$ as per Equ.~\ref{equation_t_cd}~\&~\ref{equation_t_td} for the current hour boundary. At $t_{cd}$ the monitor module retrieves the current spot price(P). If $P$ exceeds $A_{bid}$, it generates $E_{ckpt}$ event for the Controller module. On receiving $E_{ckpt}$ event, the Controller module executes $W_{ckpt}$ workflow. The $W_{ckpt}$ workflow just saves the results(the completed tasks) to the EBS volume. The Monitor module also retrieves the current spot price(P) at $t_{td}$. If $P$ still exceeds $A_{bid}$, it generates $E_{terminate}$ event for the Controller module. On receiving $E_{terminate}$ event, the Controller module executes $W_{terminate}$ workflow. The $W_{terminate}$ workflow terminates the spot instance forcefully. The Monitor module repeats the above procedure till $P$ does not exceed $A_{bid}$ at $t_{td}$ for all the subsequent hour boundaries.

If the instance is terminated at some $t_{td}$, the Monitor module will have to query for the current spot price to determine the next Available period(refer to Fig.~\ref{fig_availability}) at some specific instance of time(t*). However, the frequency of making the query is defined by the end user which may affect the job completion time slightly. At the start of the new available duration, the Monitor module generates $E_{launch}$ event for the Controller module. On receiving $E_{launch}$ event, the Controller module executes $W_{lauch}$ workflow. The $W_{launch}$ workflow launches a new spot instance as specified in $r_1$, mount the existing EBS volume to that instance and resume the remaining tasks of the job.

\section{Implementation and Evaluation}\label{sec_implementation}
In this section we analyze and compare our proposed ACC checkpointing scheme with the existing checkpointing schemes. The experiments have been carried out on 64 spot instance types of Amazon EC2 those have also been used in \cite{spot_ckpt}. The metrics used for this purpose include $job~completion~time$, $total~monetary~cost$ and the $product~of~monetary~cost$~x~$completion~time$ as the basis for comparison.

\subsection{Simulation Setup}
We have simulated the checkpointing schemes, discussed in section
\ref{sec_ckpt} \& \ref{sec_ckpt_acc}, using the same data set, parameters, algorithms
and assumptions used in \cite{spot_ckpt}. We have downloaded the simulator \cite{simulator} and applied the following modifications for our simulation setup:
\begin{itemize}
\item Modification is applied to all the checkpointing functions to rectify their \cite{spot_ckpt} wrong assumption that Amazon charges each hour by
the last price. The modified algorithm charges a spot instance by the cost of it's instance type at the beginning of an instance-hour as specified in the characteristics of Amazon EC2 spot instances.
\item A function is added to simulate the ACC checkpointing scheme discussed in Section~\ref{subsection_acc}.
\end{itemize}

In this paper we have not simulated the algorithm for determining $A_{bid}$ and $instance\_type$. Instead we have simulated the checkpointing schemes on all the 64 instance types under different $A_{bid}$ values from \$0 to \$2 with a granularity of \$0.001.

\subsection{Results and Discussion}
We obtain the simulation result for $job~completion~time$, $total~monetary~cost$ and the $product~of~monetary~cost$~x~$completion~time$ for all the EC2 instance types. To simplify the discussion, we present the result of a linux based extra large (m1.xlarge) instance type
in the eu-west-1 region. We concentrate on the performance of our proposed ACC checkpointing scheme compared to the optimal checkpointing scheme, OPT. We also include NONE, HOUR, EDGE and ADAPT checkpointing schemes in our result for completeness.
\begin{figure}[h!]
\centering
\includegraphics[width=3.5in]{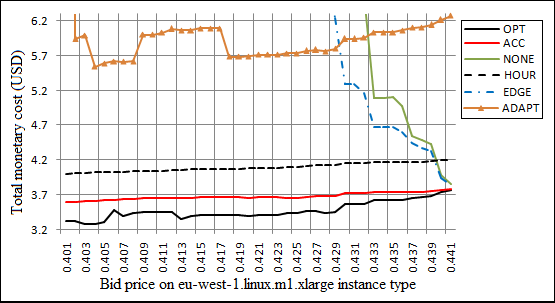}
\caption{Total monetary cost of Job completion}
\label{fig_spot2}
\end{figure}

Fig \ref{fig_spot2} shows the comparison
of total monetary cost needed to complete a job of length 500 minutes under different
user's bid($A_{bid}$) from \$0.401 to \$0.441. The result shows that ACC reduces the job completion cost significantly over the other realistic checkpointing schemes. However the cost is increased by 5.94\% on average(min 0.33\%, max 10.30\%) compared to OPT scheme. This is because the OPT scheme can execute some fraction of the job free of cost for the partial hours.

\begin{figure}[h!]
\centering
\includegraphics[width=3.5in]{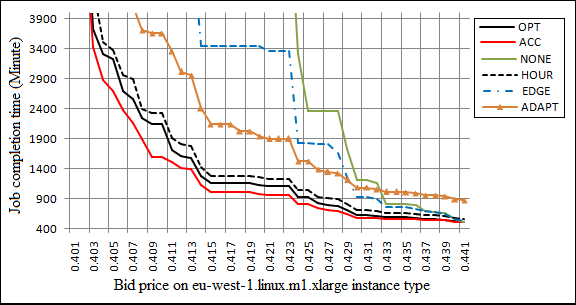}
\caption{Job completion time} \label{fig_spot3}
\end{figure}

In Fig. \ref{fig_spot3} we illustrate the comparison of various checkpointing schemes for the metric $job\ completion\ time$. Here we observe that ACC scheme outperforms all the checkpointing schemes including OPT. This is because ACC allows the job to continue even when the current spot price exceeds $A_{bid}$~(~in between a $t_{td}$ and the corresponding hour boundary~) without affecting the job completion cost. The ACC scheme reduces the job completion time by an average value of 10.77\% over the OPT scheme.

\begin{figure}[h!]
\centering
\includegraphics[width=3.5in]{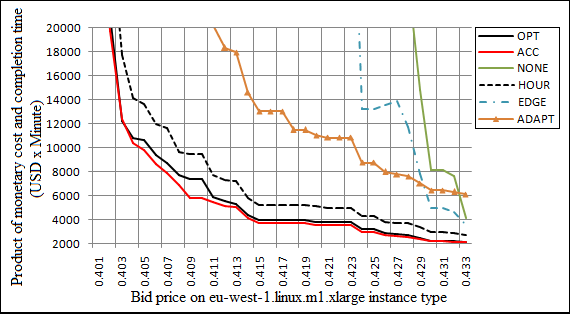}
\caption{Product of total cost and completion time} \label{fig_spot4}
\end{figure}

We plot the comparative study for the $product~of~monetary~cost$~x~$completion~time$ in Fig.~\ref{fig_spot4}. Here also we observe that the ACC scheme reduce this metric by an average value of 5.56\% over the OPT scheme.

\begin{figure}[h!]
\centering
\includegraphics[width=3.5in]{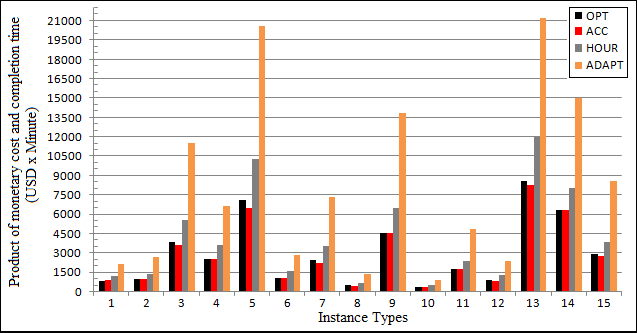}
\caption{Product of cost and completion time for different instance types} \label{fig_spot5}
\end{figure}

To gain confidence in our result, we have computed the average values of the above mentioned metrics for different bid values on all the 64 instance types. A sample of 15 difference instance types for the metric $product~of~monetary~cost$~x~$completion~time$ is shown in Fig.~\ref{fig_spot5}. For these 15 instance types, a gain of 4.03\% for ACC over OPT is observed. We also observe that such percentage gain is increased for costly instance types.

In the previous research work \cite{spot_ckpt}, the authors conclude that OPT is the optimal checkpointing scheme and none of the practical schemes can perform better than OPT. That is true only if we use the same bid values for launching the spot instance and computing the checkpoint. However our proposed ACC checkpointing scheme perform very close to OPT or even better than OPT by separating these two bid values. Thus ACC outperforms all the existing checkpointing schemes for spot instances till date.

\section{Conclusion and Future Work}\label{sec_conclusion}
Checkpointing plays an important role in reliability of job execution over EC2 spot instances. In this paper we propose a checkpointing scheme on top of application-centric resource provisioning framework that not only increase the reliability but also reduces the cost significantly over the existing checkpointing schemes. The job completion cost under the proposed scheme is very close to the optimal checkpointing scheme. Even it performs better than the optimal scheme from the point of view of job completion time, as well as product of job completion time and cost.

In future we want to investigate more on the following issues:
\begin{itemize}
\item What is the optimal bid and the corresponding instance type for a given job?
\item Should we migrate to another instance type during unavailable period?
\item What should be the new bid and the corresponding instance type for the migration?
\end{itemize}

\bibliographystyle{IEEEtran}

\end{document}